\newcommand\apj{{ApJ}}%
\newcommand\apjl{{ApJ}}%
\newcommand\apjs{{ApJS}}%
\newcommand\aap{{A\&A}}%
\newcommand\mnras{{MNRAS}}%
\def\simgt{\lower.5ex\hbox{$\; \buildrel > \over \sim \;$}}
\def\simlt{\lower.5ex\hbox{$\; \buildrel < \over \sim \;$}}

\newcommand{\msun}{\ensuremath{\, {M}_\odot}}
\newcommand{\Msun}{\ensuremath{\, {M}_\odot}}
\newcommand{\ocen}{$\omega$~Cen}

\newcommand{\Min}{M$_{\rm in}$}
\newcommand{\Mfin}{M$_{\rm fin}$}
\newcommand{\dt}{$\Delta{\rm{t}}$}
\newcommand{\ys}{Y$_{\rm s}$}
\newcommand{\yacc}{Y$_{\rm acc}$}
\newcommand{\yc}{Y$_{\rm c}$}

\documentclass[useAMS,usenatbib]{mn2e}
\usepackage{epsfig}

\usepackage[usenames,dvipsnames]{color}

\title[Pre-main sequence accretion and Globular Clusters]{Pre--main sequence accretion and the formation of multiple populations in Globular Clusters}
\author[D'Antona et al.]{Francesca D'Antona$^{1}$\thanks{E-mail:
    franca.dantona@gmail.com (FD);  paolo.ventura@oa-roma.inaf.it (PV);
    thibaut.decressin@oa-roma.inaf.it (TD); evesperi@indiana.edu (EV); annibale.dercole@bo.astro.it (AD)},
  Paolo Ventura$^{1}$,
Thibaut Decressin$^{1}$, 
Enrico Vesperini$^{2}$, \& 
\newauthor Annibale D'Ercole$^{3}$\\
$^{1}$INAF- Osservatorio Astronomico di Roma, via di Frascati 33, I-00040 Monteporzio (Italy) \\ 
$^{2}$ Department of Astronomy, Indiana University, Bloomington, IN (USA)\\
$^{3}$INAF- Osservatorio Astronomico di Bologna, via Ranzani 1, I-40127 Bologna (Italy)\\
}
\begin{document}

\date{Accepted . Received ; in original form }

\pagerange{\pageref{firstpage}--\pageref{lastpage}} \pubyear{2014}

\maketitle

\label{firstpage}

\begin{abstract}
We investigate the viability of a model in which the chemical anomalies among Globular Cluster stars are due to accretion of gas onto the protostellar discs of low mass stars. This model has been suggested as a way to reduce the large initial cluster masses required by other models for the formation of multiple stellar generations. We numerically follow the evolution of the accreting stars, and we show that the structure of the seed star does not remain fully convective for the whole duration of the accretion phase. Stellar populations showing discrete abundances of helium in the core, that seem to be present in some clusters, might be formed with this mechanism only if accretion occurs before the core of the stars become radiative (within 2-3~Myr) or if a thermohaline instability is triggered, to achieve full mixing after the accretion phase ends. We also show that 
the lithium abundances in accreted structures may vary by orders of magnitude in equal masses obtained by accreting different masses.
In addition, the same thermohaline mixing which could provide a homogeneous helium distribution down to the stellar center, would destroy any lithium surviving in the envelope, so that both helium homogeneity and lithium survival require that the accretion phase be limited to the first couple of million years of the cluster evolution.
Such a short accretion phase strongly reduces the amount of processed matter available, and reintroduces the requirement of an extremely large initial mass for the protocluster.
\end{abstract}

\begin{keywords}
stars: evolution --- stars: pre--main--sequence --- stars: abundances --- globular clusters: general --- globular clusters: individual: NGC2808, w~Cen, NGC104
\end{keywords}

\section{Introduction}
Several models have been proposed to understand the origin of multiple populations in Globular Clusters (GCs), among which the most developed are the ``AGB scenario'', in which the site of processing are the hot bottom burning envelopes of massive asymptotic giant branch (AGB) stars \citep{ventura2001, dc2004, karakas2006, dercole2008, bekki2011}, and the  ``fast rotating massive stars (FRMS) scenario''
\citep{prantzos2006, meynet2006, decressin2007a}, in which the site is the interior of fast rotating massive stars. Other interesting models have been proposed \citep{marcolini2009, demink2009, sills2010, sanchez2012, vanbeveren2012}, but they need further efforts to put them on a more quantitative ground and explore their full viability. Most of these models\footnote{An interesting different approach comes from the scenario by \cite{marcolini2009} and \cite{sanchez2012}, where the chemical anomalies are already in place in the ambient where the cluster forms.} are based on star formation of a second generation from the ejecta of stars belonging to the cluster first generation stars, and require that the initial cluster mass had to be larger (by factors estimated in 5--20 times) than the present cluster mass.

\cite{bastian2013} have recently proposed a new model according to which second generation stars are in fact born in the unique burst that gave origin to the cluster, and have modified their chemistry by accreting nuclearly processed matter lost by massive interacting binaries, during the first phases of the cluster lifetime. This model is an interesting variation of the accretion scenario proposed by \cite{dantona1983}, using the large dimensions of the protostellar accretion discs to maximize accretion and make it possible for the entire lifetime of the disc, that may reach 10--20~Myr.

Detailed studies of the abundance patterns produced by the accretion model proposed by Bastian et al. (2013) are still lacking but, on the basis of the chemical properties of the ejecta of a binary star with a primary component of 20$M_{\odot}$ and a secondary component of 15$M_{\odot}$,  these authors suggest that observed spreads abundances can be achieved and, at the same time, this model would not require the cluster to be initially more massive than now.

The observed abundance patterns provide some of the key (and more complex) constraints for all models of multiple population formation, and, in any assessment of the viability of different dynamical scenarios for the origin of multiple stellar populations, particular attention is to be paid to the scenario predictions concerning these patterns.

One of the challenges to be addressed by any model for the formation of multiple stellar generations is to understand whether it is possible to obtain significant helium differences between the first stellar population (the stars that do not accrete, that preserve the Big Bang helium mass fraction of Y$\sim$0.25) and the accreting stars. In fact, there are some clusters in which the main sequence (MS) is split, where the bluest parts of the sequence can be understood only by making the assumption that its stars have  larger (and uniformly larger) helium content. The most critical of these evidences are the MSs of \ocen\ and NGC 2808 \citep{bedin2004, piotto2007}\footnote{In both these clusters, a more direct spectroscopic evidence for helium differences has also been found among red giants, by \cite{pasquini2011} (NGC~2808)  and \cite{dupree2013} (\ocen).}.

The first studies of the chemical properties produced by the early accretion model were carried out in Cassisi \& Salaris (2014) and Salaris \& Cassisi (2014).

\cite{cassisi2014} nicely showed that there are difficulties in achieving populations with high helium, and each showing the small spread of helium that can be derived by examining the blue MS in NGC 2808.
In \cite{salaris2014}, the authors applied the same technique to explore what the early accretion scenario implies for the helium vs. lithium, or oxygen and/or sodium versus lithium patterns observed in some clusters.
It is generally recognized that the lithium abundance in GC stars may provide a strong constraint for models of the formation of multiple populations, since the same nuclear processing responsible for the chemical signatures of multiple populations fully destroys lithium in all possible models, apart from the the AGB - super--AGB scenario \citep{ventura2010}. The relation between the abundance of lithium and that of other elements varying in multiple stellar populations may provide a key fingerprint of the polluters.
\cite{dantona2012} explored the AGB-- super AGB scenario, by computing the chemical evolution in some clusters, with interesting results. In particular, they can explain the lithium -- oxygen data by \cite{shen2010}, that constitute a challenge for all models, and in particular for the early accretion scenario, as pointed by \cite{salaris2014}.

As for the evolution of lithium abundance, we remark that the \cite{bastian2013} accretion model has to deal with the problem that lithium burns as soon as the temperature in pre main sequence  reaches $\sim 2.5 \times 10^6$~K. Lithium depletion generally occurs at an age of $\sim$2--3~Myr, at the border between the radiative core and the convective envelope. So the two problems, namely the (at least partial) survival of lithium and how large the core helium content can be, are subtly connected. A possible solution is indicated by \cite{salaris2014}, and it consists in achieving the whole accretion in the first million year of the cluster evolution.

If the whole accretion process is to take place within the first million years of the cluster evolution, the amount of processed gas available will be much smaller (in fact, according to Fig.~4 of Bastian et al. (2013) no gas at all would be available before about 2~Myr) raising again the ``mass-budget problem" and re--introducing the requirement of an initial cluster mass significantly larger than the current one.

\cite{cassisi2014} and \cite{salaris2014} did not compute explicitly the accretion phase, and directly assumed that the required chemical composition could be achieved by homogeneously mixing helium rich matter with the original helium normal material of the seed star down to the center of the star, for the whole 10$^7$~yr of the disc survival around the pre-main sequence star. Thus in our computation we maintain the hypothesis that the accretion phase may last for 10--20~Myr and take a closer, although still very simplified, look at the resulting models.

In this work, we illustrate the additional problems  faced by the accretion model, by computing explicitly the accretion phase.
Although we make extreme (favorable for the scenario) assumptions on the composition of the accreting matter, assumed to have uniformly constant and very high helium content, we show that the phase during which the accreting matter can be convected to the central regions actually lasts a few million years, so that a prolonged accretion phase is not able to form stars having in the core the high and uniform helium content of NGC 2808 blue MSs. In addition, we show how the lithium surface content resulting from the accretion phase prolonged for 10~Myr, for a resulting mass of 0.7~\msun, varies by order of magnitudes, depending on the accreted mass. 

In Sect.~2 we remind why we need full mixing of the accreted matter, as the helium differences among the different stellar populations must be achieved in the stellar core to have the required evolutionary effects. Sect.~3 describes the models. The results are in Sect.~4. In Sect.~4.4 we notice that refinement of our models would lead to worse problems for the accretion scenario. In fact, we do not include thermohaline mixing, which can possibly help to fully mix the accreted matter down to the stellar core, apparently helping the accretion model for what concerns the helium enrichment. Unfortunately, if total mixing occurs (and a small molecular weight inversion should be sufficient for the onset of the mechanism), it will also fully destroy any lithium surviving in the envelope. The conclusion is that either we can not have discrete and  homogeneous contents of helium in the accreted stars, or we have to face the problem of a full lithium depletion, even in clusters with mild helium variations, contrary to present observations.

\section{Stellar structure requirements}
Observations show that most  Globular  Clusters show characteristic ``chemical anomalies'' that led to postulate two main phases of star formation. The evidence is that in many cases these anomalies are  ``mild'' ---e.g., small Oxygen depletion (by $\sim$--0.2~dex)  and relatively small Sodium enhancement (by $\sim$0.4~dex)---, in other cases the anomalies are ``strong'' ---e.g. $\simlt0.5$~dex in O and $>$+0.4~dex in Na---. In the latter case, these anomalies are also associated with very significant helium enhancement in the extreme population. While \cite{bastian2013} quote Y=0.31 for the helium content of the second generation in NGC~2808, this value describes the ``intermediate'' SG stars in this cluster, while the most extreme population apparently attains Y$\simeq$0.38 \citep{piotto2007}. Nevertheless, there can be some uncertainties in the actual figures, so it is possible to take as representative value for this population a smaller value, like Y=0.35. This is probably a reasonable value also for ``blue main sequence'' discovered in \ocen\  \citep{anderson1997, bedin2004}. The main difficulty with the accretion scenario is however the fact that the three MSs of NGC~2808, and the blue MS of \ocen\ are well separated from the others, indicating that the spread in each value of Y relevant for the population under exam is small, if any. Of course, this is the most difficult achievement to be reached by an accretion model such as the one hypothesized by \cite{bastian2013}, as shown by \cite{cassisi2014}.  

A key point in the MS location of models of different Y is the following: the structure in the phase of hydrogen burning ---the rate of hydrogen burning, the core temperature and density, and as a consequence the stellar luminosity and radius for a fixed mass and metallicity--- depend on the {\it central initial helium content, \yc}, where the burning takes place. 
The difference in \yc\ is also the key ingredient that leads to faster evolution of stars with higher \yc. The consequence is that a smaller mass is evolving in the red giant branch at the age of the GC, and the thus the stars with larger helium have smaller mass and populate hotter parts of the horizontal branch (HB) \citep{dantona2002}. This is considered an important ingredient to understand the complex HB morphology and this point is supported also by spectroscopic observations
of HBs, showing that sodium rich (second generation) are hotter than sodium normal (first generation) HB stars \citep[e.g.][]{marino2011, marino2014, gratton2012, gratton2013}.  
Thus, in exploring the viability of the accretion scenario, a key assumption is that that the star remains fully convective during the whole phase. This is the implicit assumption in \cite{bastian2013} and \cite{cassisi2014}. However,
it is important not to confuse two different timescales: the total duration of the pre--MS disc, which may channel and enhance accretion on the seed star, and the total duration of the fully convective phase. Low mass stars are convective in pre--MS for a time that depends on the stellar mass and stars with masses $\simlt$0.35~\msun\  remain fully convective down to the MS \citep[e.g.][]{dm1994}. It is then important to check what abundances can be reached in the stellar core if we follow the whole accretion phase. Lithium too burns in the pre main sequence stars at similar ages of 2-3~Myr, so we also address this complementary issue.

\begin{figure}
\centering{
\includegraphics[width=8cm]{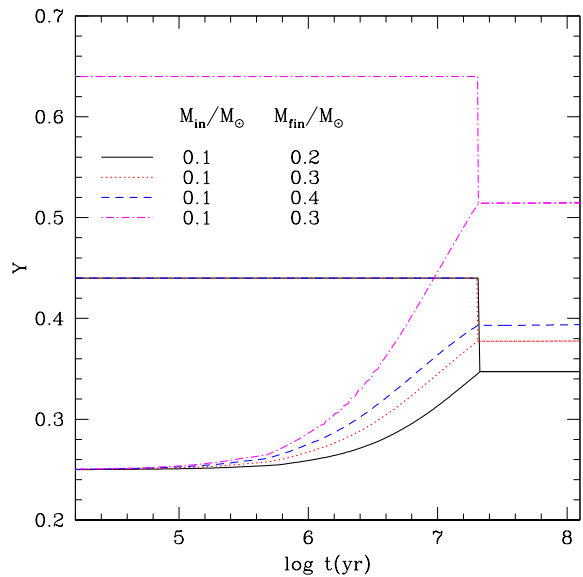}
  \caption{Temporal evolution of accreting models in which the accretion time is fixed to 20Myr. We plot the surface \ys\ and central \yc\ helium abundance as a function of log time (years), during the evolution of accreting very low mass stars. \ys\ has the value of the accreting gas until accretion stops, at 20~Myr. \yc\ increases during the whole accretion phase, thanks to full mixing of the accreting gas within the whole star. When accretion stops, \yc\ and \ys\ have the same full mixing value.  The accreting helium abundance is Y=0.44 for three cases in which the starting mass is 0.1~\msun, and final mass is 0.2 (full black line) 0.3\msun (dotted red line) and 0.4~\msun (dashed blue line). The three lines representing \ys\ are superimposed at the ordinate y=0.44 until accretion stops. One case having Y=0.64 in the accreting gas is also shown for the evolution up to 0.3~\msun, as dash-dotted magenta line.}
  }  
\label{figura1}   
\end{figure}

\section[]{Description of the models}
We use our stellar code ATON, whose recent updates are described in Ventura et al. 2008, to produce standard models of stellar masses M$\geq$0.4\msun, starting from the pre--main sequence and evolved during the entire hydrogen burning phase, for the chemical composition Z=2$\times 10^{-3}$\ and Y=0.25, 0.30 and 0.35. Additional models from 0.25 to 0.7\msun\ are also built for Y=0.37. Isochrones from these models are representative of the stellar populations present in the cluster NGC 2808 
according to the data presented in Piotto et al. 2007.

\begin{figure*}
\centering{
\includegraphics[width=5.8cm]{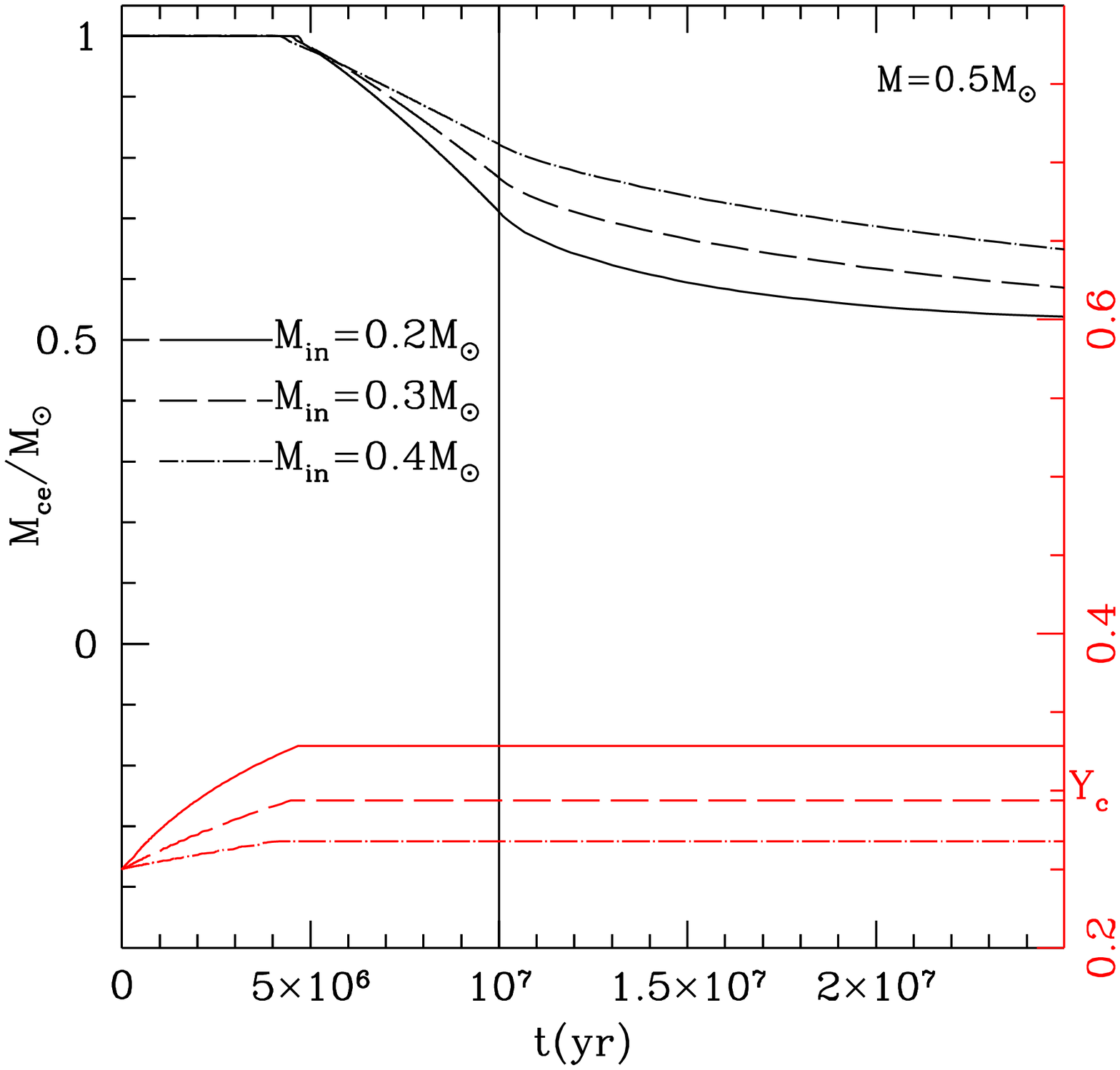}
\includegraphics[width=5.8cm]{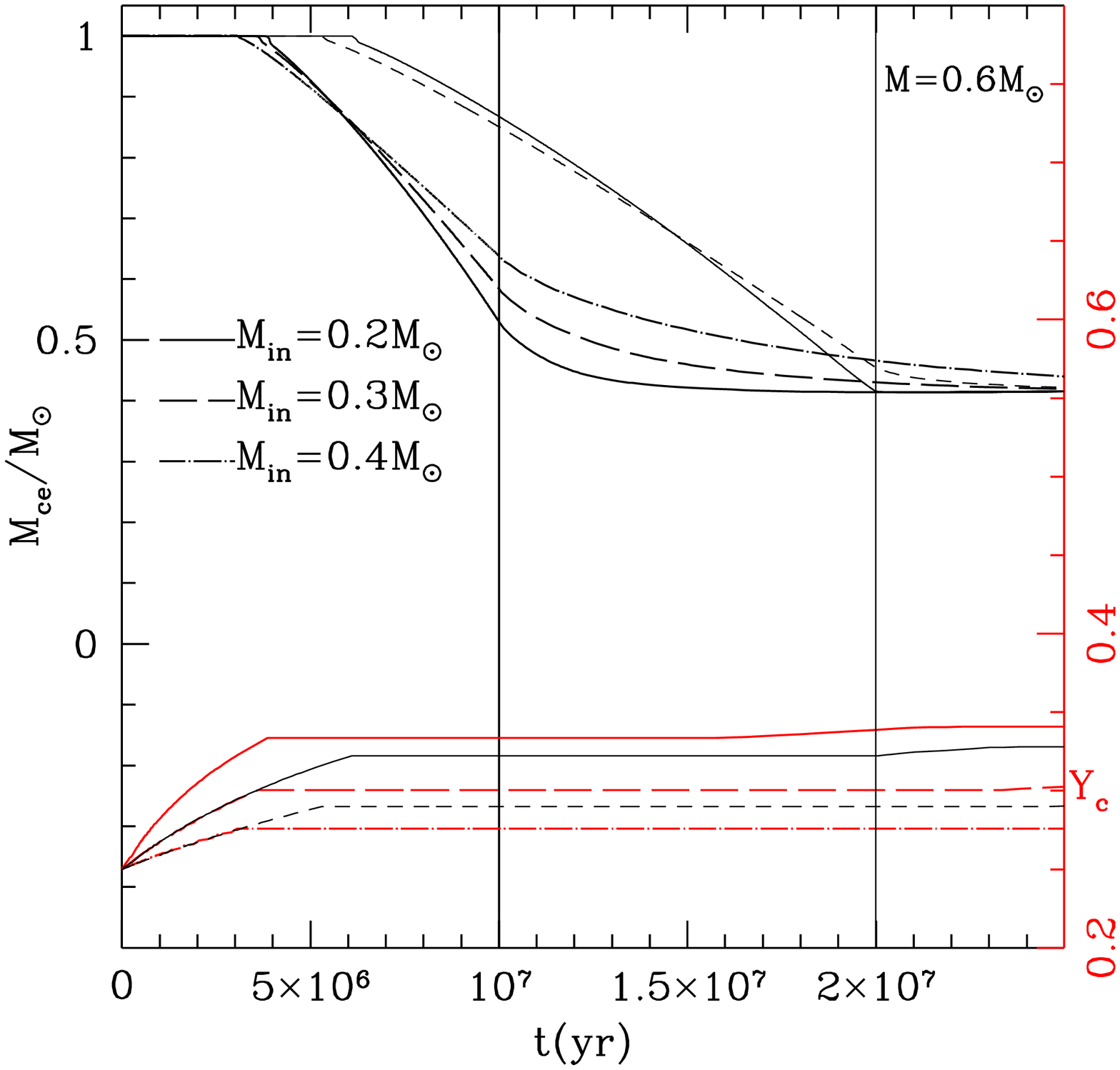}
\includegraphics[width=5.8cm]{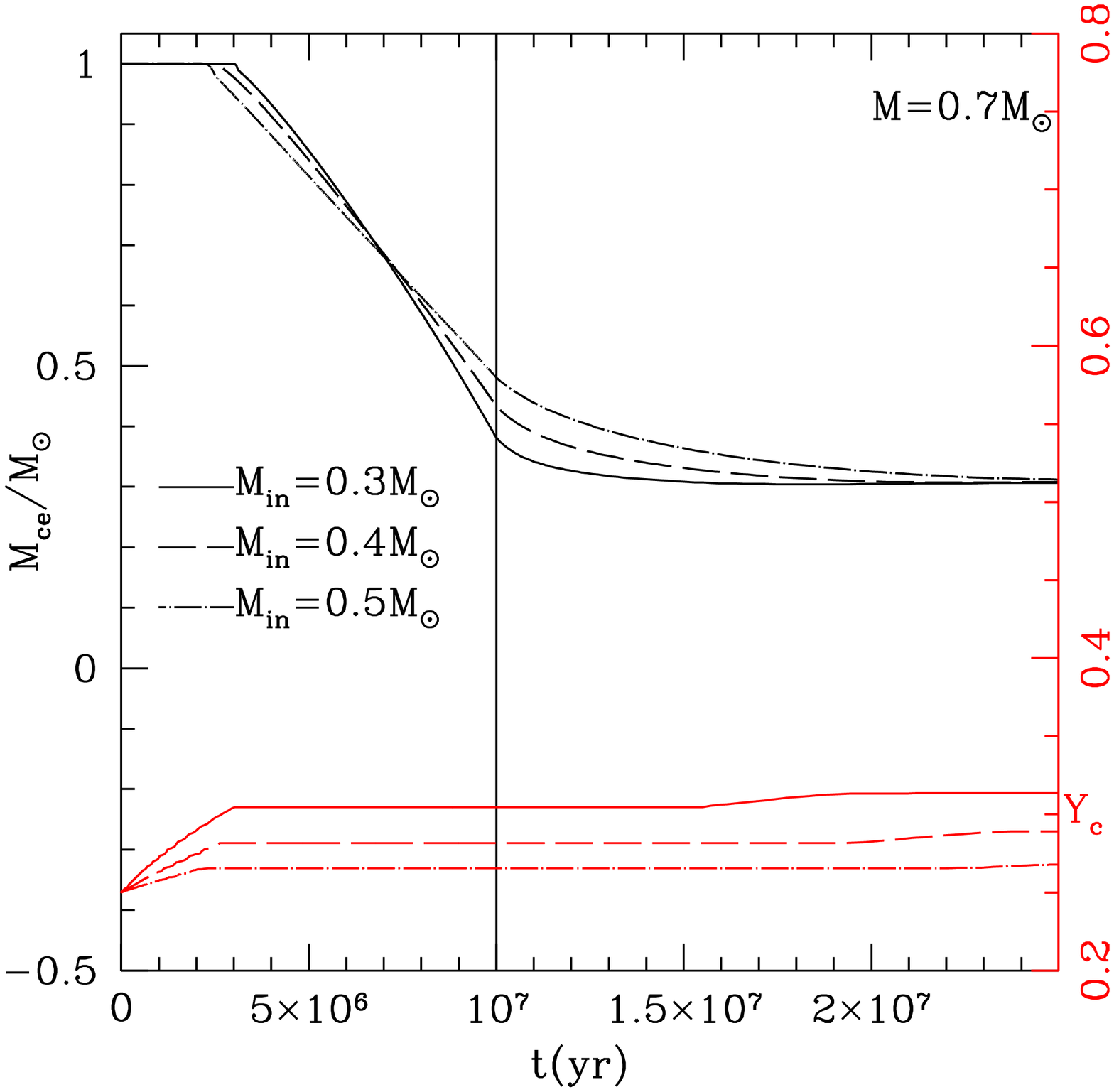}
}
  \caption{We show the evolution with time of the convective mass fraction and of the central helium abundance \yc, for three different total masses (0.5\Msun, left panel; 0.6\Msun, central panel; 0.7~\Msun, right panel) starting accretion from different seed masses, as labelled in the figures. The accreting helium content is Y=0.44, and the accretion time is \dt\ is 10~Myr for all the evolutions. Two cases for \dt=20~Myr are also shown in the central panel (here the starting masses are 0.2 and 0.3\msun).}
  \label{f2}
\end{figure*}

\begin{figure*}
\centering{
\includegraphics[width=5.8cm]{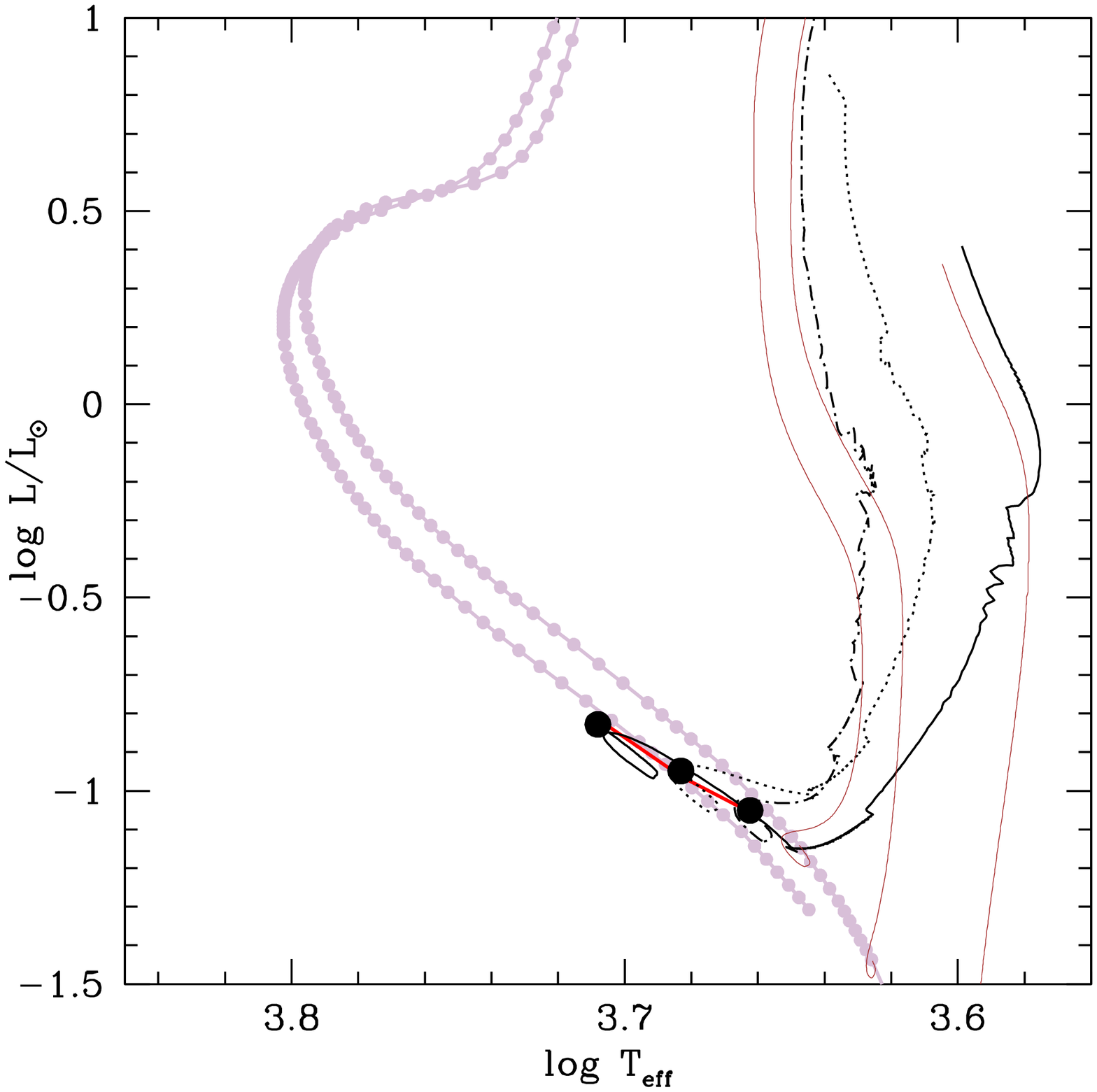}
\includegraphics[width=5.8cm]{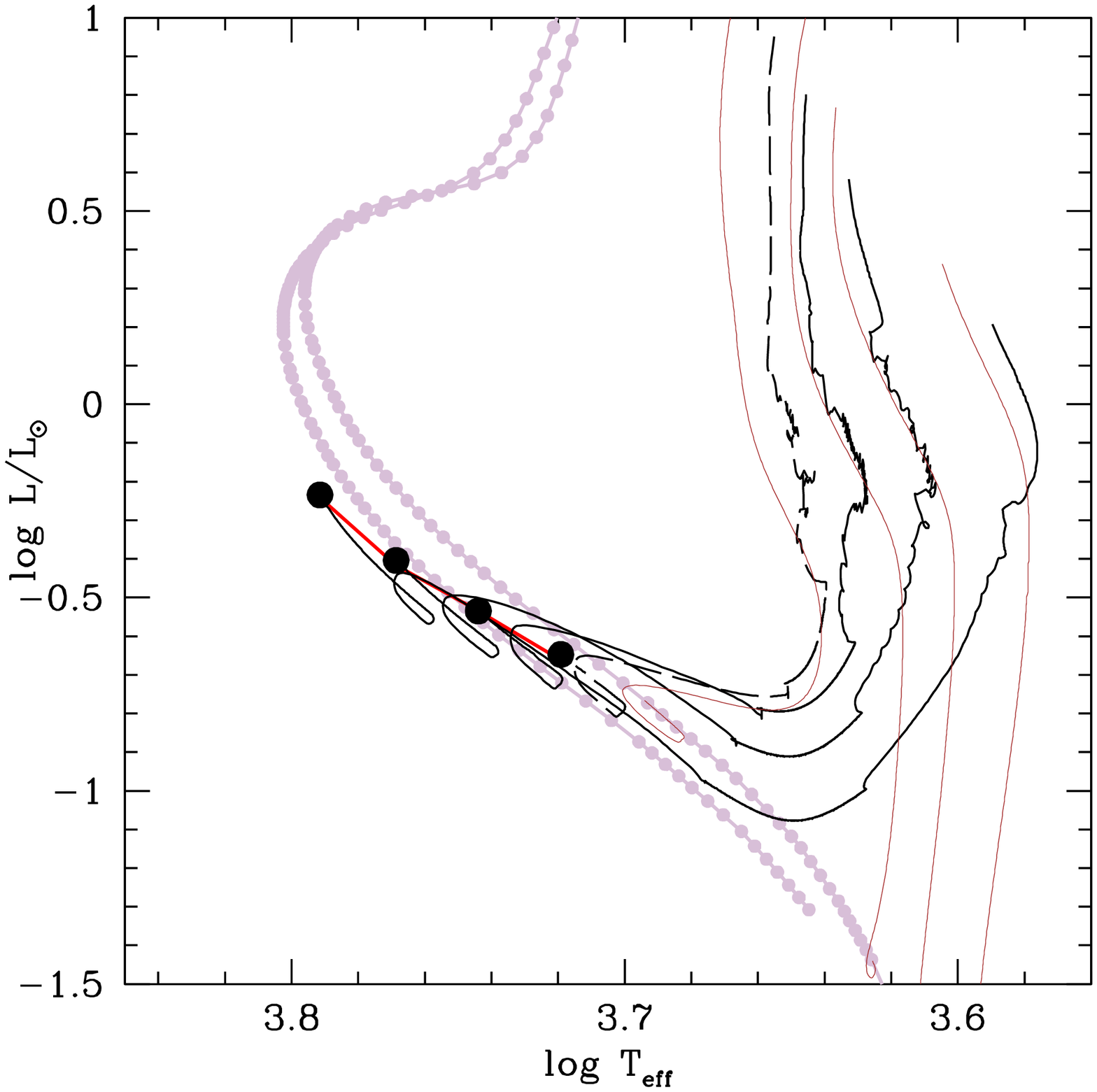}
\includegraphics[width=5.8cm]{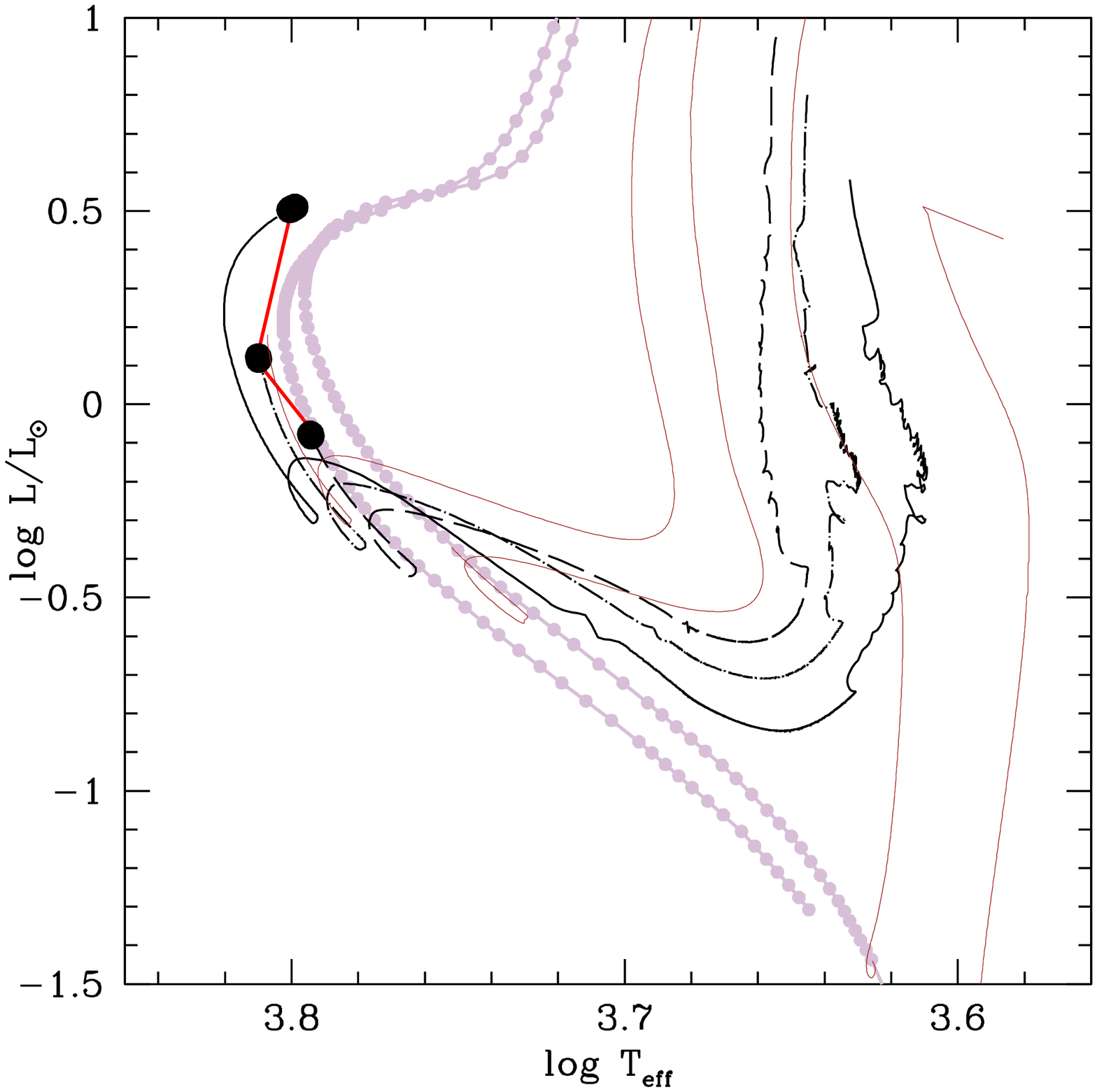}
}
  \caption{We show the evolutionary tracks in the HR diagram of the models accreting matter with Y=0.44. As a guideline we show as light full lines (in brown) some evolutions at constant mass for Y=0.25. The grey dots mark the location of isochrones of 12~Gyr for the same metallicity and for Y=0.35 (left) and Y=0.25 (right). 
All the tracks are from the \dt=10~Myr computation. Once the final mass is reached, the evolution at constant mass is shown, until the age is equal to 12~Gyr.  The red segments connect the different models at 12~Gyr, to show their location, compared to isochrones of constant helium abundance.  Left panel: evolutions ending in total mass of 0.5\Msun\ and starting from 0.2, 0.3 and 0.4~\Msun\ (locations from right to left). From right to left, the brown lines are the evolution at constant mass for 0.2, 0.4 and 0.5\msun\ (Y=0.25). Central panel: evolutions ending in a total mass of 0.6~\Msun, and starting from 0.2, 0.3, 0.4 and 0.5~\Msun; constant mass evolutions are given for 0.2, 0.3, 0.4, 0.5 and 0.6\msun\ in brown. Right panel: evolutions ending in 0.7~\Msun\ starting from 0.3, 0.4 and 0.5~\Msun. Constant mass tracks for 0.2, 0.4, 0.6 and 0.7\msun\ are plotted in brown.
}
  \label{f3}
\end{figure*}

\begin{table}
 \centering
 \begin{minipage}{140mm}
  \caption{Models}
  \begin{tabular}{lcc c c | cc}
  \hline
   \multicolumn{2}{c}{$\Delta$t=10Myr} & & \multicolumn{2}{c}{Y$_{\rm acc}$=0.44} &   \multicolumn{2}{c}{$Y_{\rm acc}$=0.64}  \\
\hline  
N     & M$_{\rm in}$  &  M$_{\rm fin}$ & Y$_{\rm c}$       &  Y$_{\rm s}$  &  Y$_{\rm c}$  &  Y$_{\rm s}$    \\
1     &0.2    &  0.3  & \multicolumn{2}{c}{0.313}  &   0.363 & 0.384 \\
2      & 0.2  &   0.4   & 0.328         &0.350  &  0.383  & 0.470 \\
3      & 0.3   &  0.4  & 0.288 & 0.298  &  0.316  & 0.356  \\     \\
4      &0.2   &   0.5  & 0.336 & 0.377  &  0.395 & 0.544 \\
  5      &0.3   &   0.5  & 0.300 & 0.366   & 0.336 & 0.438 \\
6      & 0.4   &   0.5  & 0.273 & 0.294 & 0.285  & 0.341  \\  \\
7     &0.2    &  0.6  & 0.344  &0.403  & 0.416  & 0.635  \\
8     &0.3     & 0.6  & 0.311  &0.368  & 0.349 &  0.526 \\
9     &0.4    &  0.6  & 0.284  &0.331  & 0.302 &  0.428 \\
10      &0.5    &  0.6 &  0.264  &0.291 & 0.269  & 0.337 \\  \\
11     &0.3    &  0.7  & 0.317 & 0.399  & 0.357  & 0.638  \\
12    &0.4    &  0.7   &0.294  &0.368   &  0.312  & 0.544 \\
13    &0.5    &  0.7   &0.271  &0.334   &  0.282  & 0.436  \\ \\
14    &0.4    &  0.8   &0.299  &0.415   & 0.320  &  0.639 \\
15    &0.5    &  0.8   &0.282  &0.383    & 0.290  & 0.617  \\
\hline
 \multicolumn{2}{c}{$\Delta$t=20Myr} & & \multicolumn{2}{c}{Y$_{acc}$=0.44} &   \multicolumn{2}{c}{$Y_{acc}$=0.64}  \\
\hline  
N     & M$_{\rm in}$  &  M$_{\rm fin}$ & Y$_{\rm c}$       &  Y$_{\rm s}$  &  Y$_{\rm c}$  &  Y$_{\rm s}$    \\
16     &0.1    &  0.2  & \multicolumn{2}{c}{0.347}   & \multicolumn{2}{c}{0.445}   \\
17    &0.1    &  0.3  & \multicolumn{2}{c}{0.377}   &  \multicolumn{2}{c}{0.514}    \\
18   &0.1    &  0.4  & \multicolumn{2}{c}{0.393}   &   0.538 & 0.550  \\
19    &0.2   &  0.4   &  0.312 &0.399  & 0.363  &  0.489 \\
20    &0.2   &  0.5   &  0.317 &0.390  & 0.376  &  0.610 \\
21    &0.2    &  0.6  &  0.322 &0.429  & 0.384  &  0.640 \\
22    &0.3    &  0.6  &  0.290 &0.397  & 0.336  &  0.593 \\
23    &0.2    &  0.8  &  0.329 &0.439  & 0.396  &  0.640 \\
24    &0.4    &  0.8  &  0.276 &0.438  & 0.303  &  0.640 \\
\end{tabular}
\end{minipage}
\label{table1}
\end{table}

We start from an initial   ``seed" mass \Min\ and accrete at constant rate mass with the chosen ``polluted" chemical composition. The matter is simply added at the stellar surface, at the same entropy of the outermost layers. Mixing of the accreted layer follows in the whole convective region. We adopt a simple scheme for the accretion rate, namely, we fix the final total mass \Mfin\ that we wish to achieve, and the total timescale of accretion $\Delta t$. The rate is the resulting
$$\dot{\rm M}={{\rm{M_{fin}}-\rm{M_{in}}} \over {\Delta t}}$$
We chose two possible values of  \dt: 20~Myr and 10~Myr. The longer timescale is a typical upper value to the duration of the accretion disc in star forming regions, and is the value adopted by \cite{bastian2013} for the model they propose, while the shorter time is chosen to achieve the maximum helium pollution at the center of accreting stars (see later). Actually, the disk fraction in low mass stars is about 50\% for star forming regions dated 1--3~Myr, while it is already as low as 25\% in Upper Sco, dated at 11~Myr \citep{luhman2012}. In the accretion model, the working hypothesis  is that the presence of the matter lost by the massive binaries in the cluster core extends  the activity of the disc, and its possibility to collect and accrete gas.
The evolution of \Mfin\ is then continued at constant mass until the age reaches typical age of GC stars (10--12~Gyr).
The computed models are listed in Table 1. 
We use a very high helium content in the accreting gas. Specifically we have adopted Y=0.44 and Y=0.64. The first value (Y=0.44) is the maximum value of helium in the ejecta of the massive binary simulated by \cite{demink2009}, in the first phase of mass lost to the cluster medium, while Y=0.64 is the content in the matter ejected during the second phase of contact evolution of the same binary. We emphasize that these values, then, are not average helium contents in the polluting matter, but are anyway chosen by  \cite {bastian2013} for their model.  In addition, in order to reach a final core helium content much higher than the abundance of the standard population, the model requires that the accreted mass increases  substantially the initial seed mass.

\begin{figure}
\centering{
\includegraphics[width=8cm]{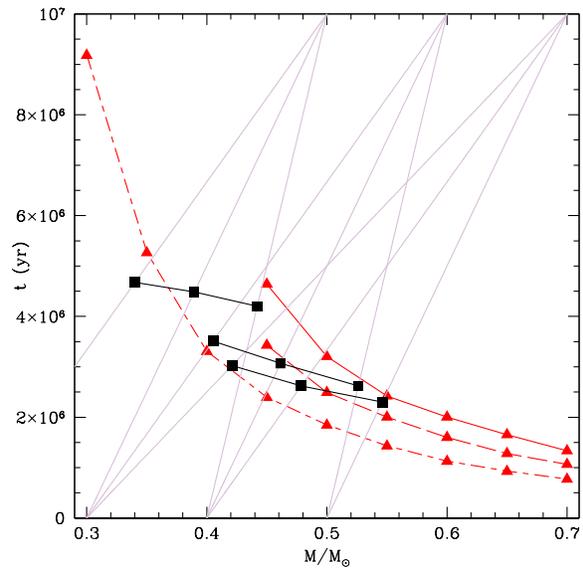}
}
  \caption{ The triangles mark the end of the fully convective phase in stars evolving at constant mass, having Y=0.25 (solid line), Y=0.30 (dashed) and Y=0.37 (dash--dotted). Note that below 0.45\msun\ the models with Y$\le$0.30 remain fully convective down to the main sequence, while the models for Y=0.37 attain a radiative core, at larger ages, down to 0.3\msun.  The diagonal grey lines show the time evolution of the mass in accreting models, for the case \yacc=0.44, ending at 0.5, 0.6 and 0.7\msun. The squares along these lines fix the mass and age at which the radiative core appears. The corresponding value of \yc\ can be read from column 4 of Table 1 (case \dt=10Myr). The figure shows that the ages at which the accreting models are no longer fully convective are close to the ages expected from evolution at constant mass for similar Y. }
\label{convduration}
\end{figure}

\section{Results}
\subsection{The most favorable case}
Figure 1 shows the case in which the hypothesis of full mixing during the whole accretion phase is met: we plot the helium surface and core content along the evolution of 0.1\msun, accreting up to total mass of 0.2, 0.3\msun\ or 0.4\msun, within a total time of 20Myr. The star remains fully convective during the whole accretion phase, and at the end the helium content reaches an homogeneous value, that could have been computed as simply as (0.25$\times$\Min+0.44$\times$(\Mfin--\Min))/\Mfin (for the models with accretion Y=0.44).
The result of these models are the only ones which conform to the starting hypothesis of \cite{bastian2013}, namely a full mixing throughout the seed star for the whole duration of the accretion phase.

\subsection{The standard case}
Figure \ref{f2} shows what happens in the general case, when the final mass is larger than $\sim$0.4\msun, and the models do not remain fully convective. We show the result of accretion up to a total mass of 0.5 and 0.6\Msun, obtained from seed masses of 0.2, 0.3 and 0.4\Msun,  and 0.7\Msun, reached from seed masses of  0.3, 0.4 and 0.5\Msun. On top (blue lines, scale on the left) we see the evolution with time of the convective mass fraction of the star, the the lower lines  show \yc. We see that full convection lasts between 2 and 5~Myr, so that only the mass accreted in the first 2 or 5~Myr increases \yc. The central panel shows that the duration of the fully convective phase is longer (5--6~Myr) when the adopted accretion time is 20~Myr.  This result depend on the fact that the stellar structure evolution depends on the mass reached by accretion at each time: a smaller \dt\ means a larger accreted mass (which evolves faster out of the fully convective phase) at earlier times. 

The HR diagrams in Figure \ref{f3} show the accreting tracks also after the accretion phase is over,  until a typical age of globular cluster (12~Gyr) is reached at the constant final mass. The MS locations of the different tracks obtained for the masses 0.5, 0.6 and 0.7\Msun\ are connected by red lines, and reflect the different \yc\ reached during the accretion phase. The 0.7\Msun\ figure also shows that at 12~Gyr the most helium rich model (that resulting from accretion on the seed mass 0.3\Msun, having a final \yc=0.317, is clearly evolved past the turnoff.

For completeness we show in detail, in Fig.~\ref{convduration}, the age at the end of the fully convective phase, as a function of the total mass, for three different helium contents (Y=0.25, 0.30 and 0.37) for the models evolving at constant mass. Notice that at each mass the time of full convection decreases for larger Y. In addition, while masses M$\le$0.4\msun remain fully convective for Y=0.25 and Y=0.30, the Y=0.37 models develop a radiative core even at 0.3\msun. We also show the straight lines of evolution with mass accretion, ending at 0.5, 0.6 and 0.7\msun\ (dotted). Along these lines we identify with squares the mass and age at which these accreting tracks begin to develop a radiative core. These times are reasonably close to the corresponding ones for constant mass evolution (taking into account the helium content reached in the model during accretion).

It is interesting to compare the results in Table 1 for \dt=10 Myr with the predictions by \cite{cassisi2014}, based on the hypothesis of full mixing for the entire accretion timescale. The central helium abundances in our models are systematically smaller than those obtained in their framework. For example, for the case \yacc=0.44, their 0.2 (initial) to 0.6 (final) \msun\ sequences ends in a blue main sequence star in NGC~2808, with Y$\sim$0.38, but the central abundance in our corresponding sequence is \yc=0.344. Their 0.4 (initial) to 0.7 (final) \msun\ sequence ends on the middle MS of NGC~2808, at Y=0.33, but our corresponding sequence only attains a central Y=0.294.

\subsection{Exemplifying the lithium mixing case}
\begin{figure*}
\centering{
\includegraphics[width=8cm]{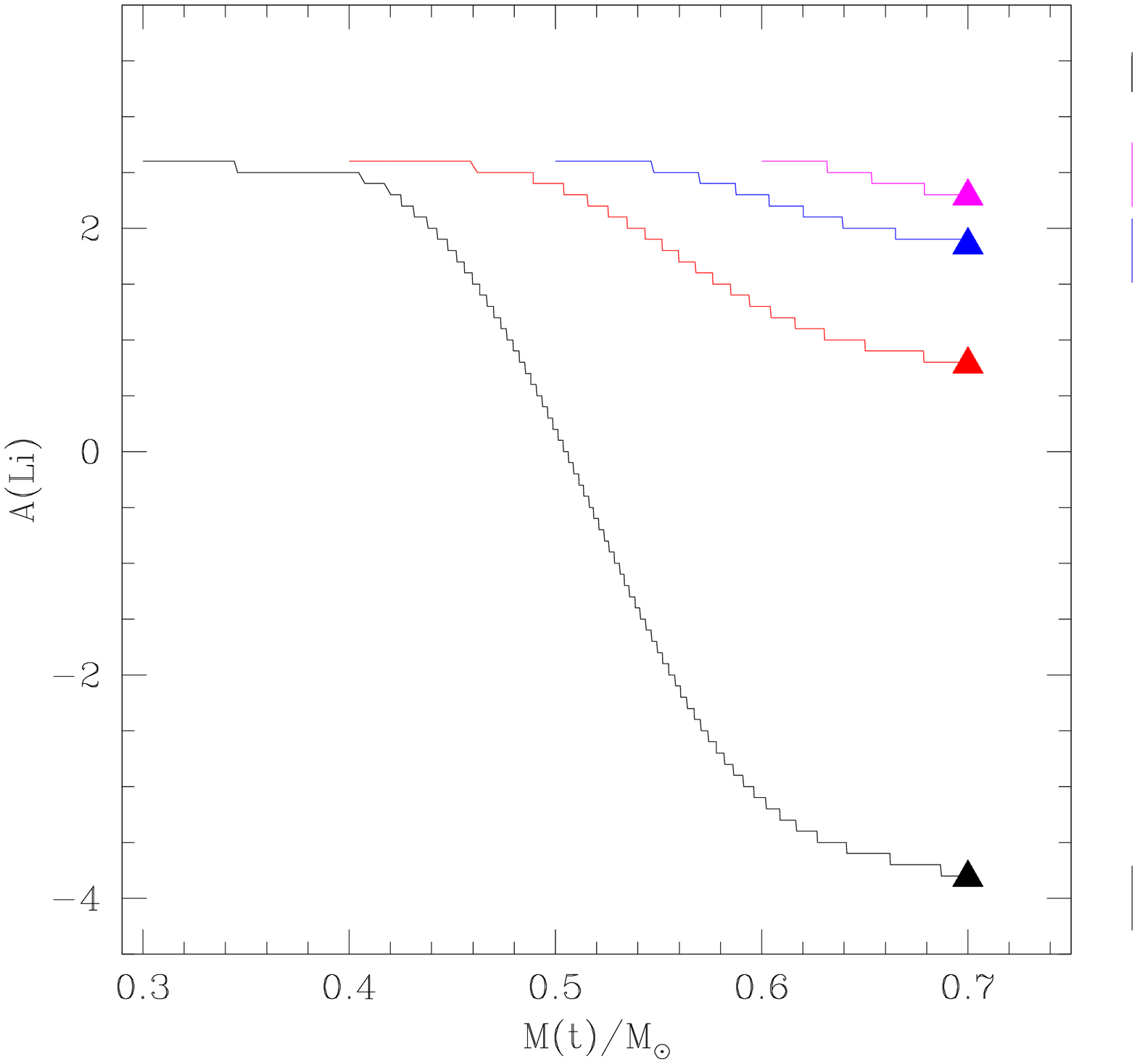}
\includegraphics[width=8cm]{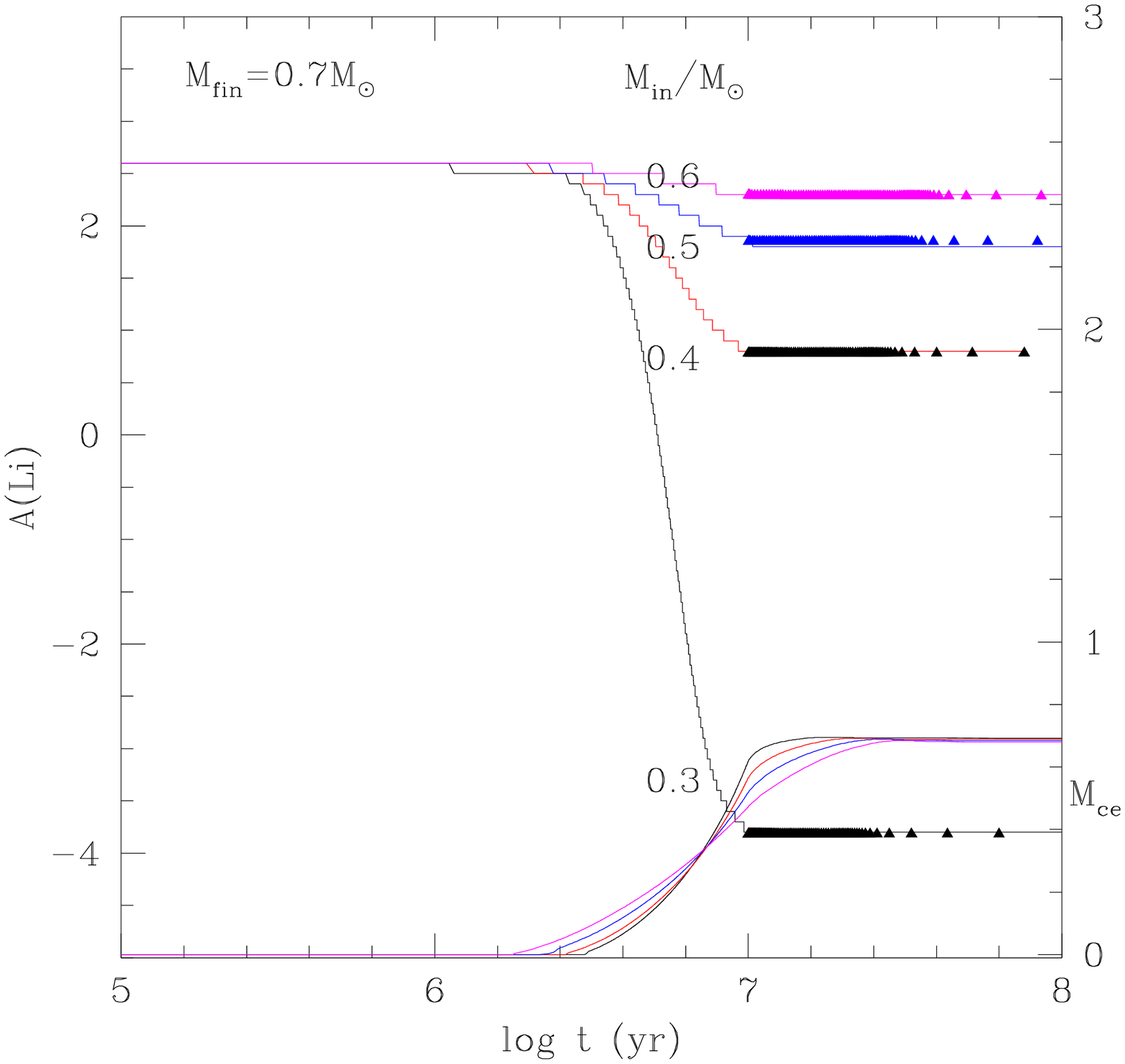}
}
  \caption{Variation of the lithium abundance at the basis of the convective envelope (lines) as a function of the time (right) and of the accreting total mass (left) for models accreting for 10$^7$~yr from initial masses 0.3, 0.4, 0.5 and 0.6~\msun, and all ending at M=0.7~\Msun.  The resulting surface lithium abundance in the final 0.7\Msun model is highlighted by triangles. The core, or the interface between the convective envelope and the radiative core (shown at the bottom of the right figure as fractionally mass M$_{\rm ce}$), reaches  the temperature for lithium burning  at an age between $\sim 1.6$\ (for the initial mass 0.3~\Msun) and $\sim$3~Myr (for the initial mass 0.6~\Msun). }
\label{lithium}
\end{figure*}

Figure \ref{lithium} shows the results of simulations in which the total mass obtained is 0.7\msun, starting from 0.3, 0.4, 0.5 and 0.6\Msun, and keeping the timescale at 10Myr. The left panel shows the temporal evolution of the lithium content (full lines) at the inner border of the convective envelope: the phase of lithium burning is achieved at an age of $< 3 \times 10^6$~yr. At the same time, as shown in the right panel of Fig. \ref{lithium} (lines at the bottom), the convective core disappears, and we see at the surface the effects of lithium burning  at the retreating border of the convective envelope\footnote{Obviously, lithium is also fully depleted below the convective envelope, as the temperatures there are larger, but the effect is not seen at the surface, unless, or until, extra--mixing mechanisms occur --- see Sect.\ref{doesnotwork}.}. The final results of the four different cases is shown by the triangles, which indicate the level of lithium abundance 
in the final 0.7~\msun\ models, at the end of the accretion phase. Obviously, the abundance is lower for larger amounts of Li-depleted accreted matter. This confirms \cite{salaris2014} result that very fast accretion is needed to maintain a non negligible lithium content in the accreting stars. 

\subsection{Possible effects of  internal mixing}
\label{doesnotwork}
Our models do not take into account mixing except in convective zones.
Chemicals elements can be transported by rotation-induced mixing,
  atomic diffusion processes and thermohaline instability. Rotation-induced
mixing and atomic diffusion will act on long evolutionary time-scale
compared to the accretion time (see e.g., \citealt{lagarde2012a} and \citealt{richard2005}). On the contrary thermohaline mixing can act
on shorter timescale. In the case of accreting polluted matter onto the
stars, the larger helium content in upper layers leads to
the inversion of molecular weight and will provoke a thermohaline instability \citep{ulrich1972, kippen1980} that could well succeed in chemically homogenise the star again. The efficiency of thermohaline mixing is still uncertain. Concerning the extra--mixing needed at the level of the red giant branch evolution in 1D models, see, e.g., \citet{charb2010,lagarde2012}, versus \citet{cantiello2010,wachlin2011}. Concerning 3D simulations see \cite{denissenkov2011, traxler2011}, versus \cite{brown2013}. In the present case, one important question is whether the homogenization of the chemistry is fast enough to lead to hydrogen burning modalities dominated by the fully mixed or by an intermediate helium stratification. Much further work is necessary to solve this issue. 

Furthermore if thermohaline instability is occurring, the upper part of the
stars will also be mixed. This will have huge impact for fragile elements like
lithium which will be efficiently destroyed when thermohaline mixing links
the surface (still lithium-rich) to the underneath lithium-burning region. Such effect
has been  investigated by  \citet{theado2009} and \citet{theado2012} in the
case of accretion of planetary (metal rich)   material to MS and PMS stars.
 Thus if the accretion takes place after the convective envelope retracts on
  the early PMS, the star will develop thermohaline instability and most of
  its lithium will be destroyed.
It is important to stress here that, if thermohaline mixing is effective, it will be able to fully mix not only those stars for which we need to achieve a great helium overabundance in the core, as it is necessary to explain extreme cases such as NGC~2808. Even stars in which the overabundance of helium in the envelope is small  will be subject to thermohaline mixing, and lithium will be fully destroyed! In this case, then, we should not see lithium in clusters like M4 or 47~Tuc, where  Na-rich stars show a very mild lithium depletion \citep{monaco2012, bonif2014}. 
  The only possibility to avoid such Li destruction will be that the accretion takes place before the radiative
  core appears. In this case, accretion will be limited to the first
  2--3~Myr of the cluster evolution. Thus only a limited number of binaries
  (or fast rotating massive stars) will be able to contribute to the
  formation of stars with abundance anomalies. In fact, according to \cite{bastian2013} (see their Fig. 4), no gas is released in the first 2 Myr.

\section{Conclusions}
In this work we have outlined two problems that must be taken into
account in the assessment of the viability of  the early accretion model of multiple populations in
GCs. Both these problems are due to the constraints posed by  pre--main
sequence evolution on the chemical composition resulting  in the accreted
star. The first problem is that the fully convective phase of stars of
M$\simgt$0.5\msun\  lasts for the first $\sim$3~Myr of evolution, so that
massive accretion, such as envisioned by \cite{bastian2013} to account for
about  half of the total stellar mass of the anomalous stars, will produce
a smaller central helium content than obtained by simply averaging the
helium of the polluting matter and the initial helium of the star. As there
is no reason to believe that a cosmic conspiracy allows exactly the mass
accretion necessary to provide similar helium contents to different masses,
the result is that illustrated by Figure \ref{f3}, where we show
that the location of the accreted stars of different masses do not follow
isochrones of different helium content, but are spread at locations such
that results as the "triple main sequence" of NGC~2808 can not be
obtained. Furthermore the stochastic process of accretion on low-mass
 stars will be difficult to reconcile with discrete sequences.

The second problem is that also the burning of lithium begins for the pre--main sequence stars when the fully convective phase ends, at the boundary of the convective envelope. Consequently, the accretion of different mass budgets, reaching the same final mass, results in very different surface lithium contents at the surface of the accreted stars. 

We notice that additional mixing mechanisms, such as thermohaline mixing, can operate to homogenize the helium content within the stars, but would also be fully destructive for lithium, contrary to the evidence of the presence of finite lithium abundances at the surface of second generation stars.

One way to avoid these problems is to require that the whole accretion phase takes place in the very first million years of the cluster life, as suggested by \cite{salaris2014}. However, very little mass or no mass at all is released in the first one--two million years, and therefore the early accretion scenario is affected by the mass budget problem (and possibly by a much more severe form of it) that it was meant to solve.

\section*{Acknowledgments}
We thank Maurizio Salaris for his careful reading of the manuscript and for a useful referee's report.
F.D., A.D and P.V. have been supported by the PRIN--INAF 2011 ``Multiple populations in globular clusters: their role in the Galaxy assembly", P.I. E. Carretta.
TD acknowledges financial support from the European Union Seventh Framework Programme (FP7/2007-2013) under  grant agreement n\textsuperscript{o} 267251 Astronomy Fellowships in Italy (AstroFIt).
EV was supported in part by grant NASA-NNX13AF45G.

\end{document}